\documentstyle[12pt]{article}

\clearpage

\begin{document}
\baselineskip 24pt
\begin{center}
{\Large\bf Phonon-Drag Thermopower at High 
Temperatures}\\[5mm]
{\bf V.A.Vdovenkov}
\end{center}
\baselineskip 16pt
\begin{center}
 Moscow State Institute of Radioengineering,
 Electronics and Automation 

(technical university)\\
 Vernadsky~ave. 78,~117454~Moscow,~Russia\\
\end{center}
\begin{abstract}
The adiabatic cristal model is offered. It is shown that springy nuclei 
oscillations relatively electronic envelops and waves of such oscillations 
(inherent oscillations and waves) may exist in crystals. The  analysis of 
experimental temperature dependencies of resistivity in semiconductors 
with electron-vibrational centres has shown that  inherent oscillations 
effectively interact with crystalline phonons as well as with electrons 
and holes, creating  powerful interaction of electrons and holes with
 phonons. The experimental narrow peaks of phonon-drag  thermoelectric 
power at Debye temperatures from 77K to 700K confirm existence of 
inherent  oscillations waves in crystals. Inherent oscillations and waves
 gives rise to strong electron-phonon interaction and probably can bring 
about superconductivity at temperatures as below so and well above room 
temperature.
\end{abstract}

\hspace{1.5em}Oscillations of cristals traditionally have always researched
by analysing motion equations of their models. Atoms in such models are
replaced by spots with masses of atoms. These models do not correspond to
an adiabatic 
theory of solids because  the nucleus and electron envelop of each atom
in these models is considered to be hard are bound. Becouse in such 
conditions exchange of the energy between nuclei and electronic envelops 
is possible, the adiabatic principle is broken. That is why traditional 
crystal models are nonadiabatic. It's possible to find out new phisical 
properties of crystals by researching their adiabatic models. 
So  substantiated  adiabatic crystal model is important to build. 

\section*{1.~Description of adiabatic model}
\hspace{1.5em} It's possible to build an adiabatic model of cristal in 
accordance whith adiabatic solution of stationary Shredinger equation:
$(T_e+T_z+V)\psi=W\psi$, where $T_e$ and $T_z$ - kinetic energy 
operators of electrons and nuclei, V - a cristalline potential, 
$\psi$ - wave function and W - energy of cristal. This equation may be 
re-arranged to two equations supposing $\psi=\phi\Phi$ and separating 
variables:
\begin{equation}
\label{hf1}
      (T_e + V)\phi = E\phi , ~~~~~~~~   
\end{equation}
\begin{equation}
\label{hf2}
      (T_z + E + A)\Phi=W\Phi,  
\end{equation}
where E~-~energy of electrons and A~-~an adiabatic potential. Wave 
function $\phi$ describes electrons motion in the cristalline potential 
field~V. Wave function~$\Phi$ describes nuclei motion. Because of presence
adiabatic potential~A in the equation~(\ref{hf2}) the functions~$\phi$ 
and~$\Phi$ are mutually dependent, therefore electrons and nuclei in the 
cristal have a possibility to exchange the energy one with another.
So the problem of cristal oscillations study on bases of 
equations~(\ref{hf1}) and~(\ref{hf2}) in general is nonadiabatic. 
However, if potential A is small and its contribution to the energy of 
cristal can be neglected, it follows the adiabatic approach of 
Born-Oppenheimer~\cite{Born27}. In this approach an exchange of the 
energy between electrons and nuclei in cristal is absent and
equation~(\ref{hf2}) has a solution regadless of equation (\ref{hf1}).
Hartree-Fock method~\cite{Hart28,Fock30} may be applied to the 
equation~(\ref{hf2}) to define an efficient potential~$V(R_j)$, which
is dependent on coorinates only of j-th nucleus, and reduce a problem
about nuclei motion to one-partical problem:
$[T_j+V(R_j)]\Phi_j~=~W_j\Phi_j$, where $\Phi_j$ - wave function 
of nuclei, $T_j$~-~a kinetic energy operator, $W_j$ - a spectrum 
of j-th nucleus stationary energy levels in the potential field~$V(R_j)$.
Potential~$V(R_j)$ is defined by all electrons and nuclei exept 
j-th nucleus.Analysis shows that the main contribution to~$V(R_j)$ 
is introduced by s-electrons of K, L and M~electronic orbits of j-th atoms. 
Minimum~$V(R_j)$ defines a position comparatively which nucleus is 
capable realize oscillatory motion. It is shown in~\cite{Davi73}  that 
adiabatic method equitable if the energy of nucleus oscillations does not 
exceed energy of electronic transitions. Under such condition the 
energy of nucleus oscillations can not be sent to electrons. So electronic
envelops stay nondeformed and moutionless but oscillatory motions of 
nuclei are principally possible. These oscillations can be called
inherent (I) oscillations~\cite{Vdov96}, since their properties are 
defined by inherent atom parameters - by mass, charge
of nucleus and potential~$V(R_j)$ near centre of electronic envelop.
Consequently, every atom of crystal in adiabatic model must be 
presented in a manner of inherent (I) oscillator which consist of 
nucleus and electronic envelop bounded with each other by springy
force.

Nuclei oscillations relatively atom envelops occur in small 
interval $(\approx10^{-2}A^0)$ and in general it is nesessary to research
them by quantum methods. But, if study harmonic oscillations
then it is sutable to use the known correspondence between results of 
quantum
and classical theories. This correspondence consists in coinciding the 
frequency of transition between quantum adjasent oscillatory levels of
harmonic oscillator with his classical oscillations frequency. So energy
spectrum of cristal harmonic oscillations is possible to reseach by clasical
method. We used this possibility for description of oscillations in
adiabatic simple univariate cristal model. Such model is shown on 
insertion fig. 1 where circles are electronic envelops of atoms
and spots in circles centres are nuclei . Lattice spasing is equal
to~{\bf a}.
 System of motion equations for this model may be 
written as:
 \begin{equation}
\label{em1}
M\frac{d^2}{d{t^2}} U'_{n} =-\theta_1(U'_{n}-U''_{n});~~~~~~~~~~~~~~~~
~~~~~~~~~~~~~~~~~~
\end{equation}
\begin{equation}
\label{em2}
m\frac{d^2}{d{t^2}} U''_{n} =-\theta_1(U''_{n}-U'_{n} )-
\theta_2(2U''_{n}-U''_{n-1}-U''_{n+1}),                                     
\end{equation}
where M~-~nucleus mass, m~-~an electronic envelop mass, $U'$ and
$U''$~-~displacings of nucleus and electronic envelops, $\theta_1$
and $\theta_2$~-~force factors, t~-~time and elementary cell number 
$n=0,~\pm1,~\pm2,~ \cdots$ . If search solutions of this system of 
equations as harmonic waves, so the dependency of angular
frequency $\omega$ from wave vector $\bf q$ can be expressed as
\begin{equation}
\label{omeg}
\omega_{1,2}({\bf q})=(G/2)\{1 \pm \sqrt{1-(4\theta_1\theta_2)/MmG^2)}\},
\end{equation}
$G=\beta/m^*+{\gamma}C/m$, 
$C=4\sin^2({\bf aq}/2)$, ${\bf a}$~-~lattice spasings, 
$m^*=(1/M+1/m)^{-1}.$
Qualitative curves  $\omega$({\bf q}) are shown on 
the fig.~1 and contain known 
acoustic~(A), as well as new~(I) branches. In complex real cristals except
known acoustic and optical branches  there are also branches of inherent
oscillations. The number of I-branches is equal to 3r (r - a number of atoms 
in the utit cell of crystal). I-oscillations and waves can exist even 
though elementary cell  kepts only one atom and optical 
oscillations are absent. 

\section*{2.~Distinctive  properties of iherent oscillations and waves}
\hspace{1.5em}  
Energy spectrum of inherent oscillations is possible to calculate 
taking into account an interaction between oscillations of different 
electronic envelops. Displacements of electronic envelops may be coherent
within certain area with the distinctive size $\Lambda$ due to the 
interaction between them. Thermodynamic analysis of adiabatic crystal 
model gives the value: $\Lambda=\sqrt {Zm^*/ne^2\mu}$, 
where Z~-~an atomic number, $m^*$~-~effective mass of electronic envelops, 
e~-~a charge of electron, n~-~electronic density, $\mu$~-~magnetic 
constant ~\cite{Vdov96}. The value~$\Lambda$ may exeed groups 
of ten lattice spasings. In such conditions a nucleus of each atom can 
realize oscillatory motion relatively big mass of crystal  coherent area. 
Analisis shows that $\alpha$, $\beta$, $\gamma$~-~types of inherent 
oscillations and waves are to exist in dependence of displacements  K and
L~-~orbits. $\alpha$-tipe of I-oscillations corresponds to nucleus 
oscillations relatively electronic envelop. Joint nucleus and K~-~orbit 
oscillations relatively remaining part of electronic envelop correspond 
to $\beta$~-~oscillations. Joint nucleus, K and L~-~orbits oscillations 
relatively remaining part of electronic envelop correspond to 
$\gamma$~-~oscillations. Elementary quantum ($\hbar\omega_z$) of 
$\alpha$~-~oscillations for neutral atom with number~$Z>2$ calculated 
with provision for shelding of nucleus by electrons can be expressed as:

\begin{equation}
\label{quv}
\hbar\omega_z=\sqrt{\frac{e^2}{3\pi\varepsilon_0(m_n+m_p)}
\left\{\left(\frac{\vartheta}{a_0}\right)\frac{Z-\zeta-\xi}{Z-\zeta}
\right\}^3\frac{\chi(Z-\xi)}{Z}}. 
\end{equation}
Shelding is taken into account by~$\zeta=5/16$ and $\xi=sZ^{1/3}$, s~ranges
from~1 to~1.15 when changing an atomic number~Z from~8 to~80, 
$\chi=1.2$~takes into account a contribution in electronic density from 
s-states of L, M and N~-~orbitals, $\vartheta=0.88534$, $a_0=0.529 A^0$,
$\varepsilon_0$~-~electrical constant, $m_n$ and $m_p$~-~masses of 
neutron and proton. Elementary quantum of $\beta$~-~oscillations possible 
to define on the eq.~(\ref{quv}), having place~$\chi=0.2$.  Elementary
quantum of $\gamma$~-~oscillations possible to define on eg.~(\ref{quv}),
having place~$\chi=0.056$. Energy spectrum of harmonic inherent $\alpha$, 
$\beta$ or $\gamma$~-~type oscillations and waves are possible 
to describe by harmonic oscillator formula
\begin{equation}
\label{harm}
E(\nu)=E_o(\nu+1/2),
\end{equation}
where oscillatory quantum number $\nu=0, 1, 2, \cdots $, and 
$E_0=\hbar\omega_z$ for oscillations $\alpha$, $\beta$ or  
$\gamma$-type.

\section*{3.~Interaction of inherent oscillations  with phonons}
\hspace{1.5em}Inherent oscillations and waves may exist in 
the ideal crystal but  generate such oscillations and waves is  
possible (to the account of electrons or holes energy) by means of local 
centres with strong electron-phonon interaction. Inherent oscillations 
and waves distort a crystal. 
They are  capable effectively interact with other springy oscillations 
as well as  with electrons and holes, ensuring efficient electron-phonon 
coupling. This can cause a phonon-drag component of thermoelectric power 
and other physical effects. So interaction of inherent oscillations with
acoustic phonons is possible to track at simple univariate crystal model 
adding in to the equation~(\ref{em2}) additional force~-$\delta U''_n$.
This force describes effect from the coherent areas.
Corresponding dependencies $\omega(\bf q)$ is possible to get by replacing 
$G$ on $(G+\delta U''_n)$ in eq.~(\ref{em2}). Acoustic branch suffers most
alteration alongside centre of Brillouin area. At the condition $\delta>0$  
a forbidden frequency area $0 \cdots \omega^*$ appears. At condition 
$\delta<0$ a forbidden area for wave vectors $0 \cdots \bf q^*$ appears. 
Dependencies $\omega(\bf q)$ for $\delta>0$ and $\delta<0$ are shown 
qualitatevely on fig.~1 by dashed qurves. Obviously at the conditions 
$\omega^*>0$ and $|q^*|>0$ experimental acoustic methods for crystal 
oscillations studies may turn out to be ineffective.

Acoustic phonons at the conditions $0<\omega<\omega^*$ or $0<|\bf q|<|\bf q^*|$
can not exist in crystals and consequently can not scatter  charge  
carriers. For this reason superconductivity is possible if other 
mechanisms of scattering are absent.  

\section*{4.~Experiments, results,  discussion}
\hspace{1.5em}It is known that at temperatures $T<70K$ thermoelectric 
power (TEP) may
contain a phonon-drag component (PDC). Absence of PDC at $T>70K$ earlier
was explained by insufficiently strong coupling between electrons and
phonons \cite{Tc70}.  Such opinion in common was saved up to present time.
However TEP in cristalline ropes of carbon nanotubes supposedly 
was explained by contribution of PDC at temperatures from~4.2K to 
300K~\cite{Hone98}. In carbon nanotube films on  substrates at
temperatures up to~600K  narrow pius of TEP are discovered, connected with 
PDC~\cite{Vdov98,Kos98}. PDC in carbon nanotube films on substrates is 
stimulated by the interaction of electrons in film with phonons of 
substrate due to inherent $\alpha$-oscillations of Carbon~($E_0=0.25$~eV)
and Oxygen~($E_0=0.22$~eV) atoms. In this work we measured PDC  in 
monocristals GaP and Si, in porous silicon on silicon substrates.

In experiments were used flat semiconductor samples with the thickness
200~mkm containing local centres with the strong electron-phonon coupling.
We investigated GaP samples with impurities of aluminum or sulphur 
($\approx10^{15}$cm$^{-3}$): GaP(Al) and GaP(S). Such impurities were 
choosen becouse atoms Al and S have masses vastly exeed mass of atom Ga.
This advantages emergence electron-vibrational centres and generation of 
inherent oscillations. Investigated silicon samples had impurities of
Phosphorus ($\approx 5\cdot 10^{15}$cm$^{-3}$) and Oxygen($\approx 
10^{18}$cm$^{-3}$): 
Si(P,O). Before measurements the samples where subjected to heating in 
vacuum within 5 minutes at~T=600K  then they were cooled up to room 
temperature within 0.2 minues. Concentration of Oxygen in Si was 
determined on the basis of data about intensities of distinctive 
optical absorbtion near 9 mkm.
Layers of porous silicon ($Si^*$) with the thickness about 0.3~mkm on 
n-silicon substrates with resistivity 3~Ohm$\cdot$ cm were also 
investigated. In samples Si(P,O) and in Si$^*$ local centres seem to be
formed by atoms of oxygen (A-centres) for which the constant of 
electron-phonon interaction is equal to~5.

Their were measured temperature dependencies of electrical resistivity 
$\rho$(T) and temperature dependencies of thermoelectric power in 
samples within the range from 77K to~700K for the reason of studying 
the influence of inherent oscillations and waves on  migration of 
charge carriers. Temperature difference at mesurements of TEP did not 
exeed $3K \pm 0.2K$.
The spectrums of infra-red reflection modification (dR) were measured, caused
by impurities in GaP samples, in the optical range from 15 mkm (83 meV) to
2~mkm (620 meV) at~300K. The angle between nonpolarized light beam and flat 
surface of sample was equal to~$45^0$.

Typical temperature dependencies of electrical resistivity 
$\rho$(T)/$\rho_0$ of GaP samples are brought on fig.~2 in
semilogarithmic coordinates.
{$\rho_0$} is a constant and does not influence magnitudes $E_a$. 
Magnitudes {$\rho_0$} are selected for each curve that the curves 
were conveniently plased on the fig. 2. Cuve 1 on fig~2 referes to 
temperature dependencies of electrical resistivity of GaP sample without 
impurities.
Typical dependencies $\rho$(T)/$\rho_0$ of GaP(Al) and GaP(S) samples 
are presented on fig.~2 by cuves~2 and~3. These curves are broken straight  
lines. Tangent to these curves select areas which correspond to concrete 
values $E_a$. Values $E_a$, measured at $T<330K$, are contributed in 
Table~1 where stars choose the lines with energies fit
known energies of phonons in GaP~\cite{Mar64} which are
brought here in the central column. Activation energies of GaP(Al) and
GaP(S) measured at $T>330K$ are contributed in Table~2 where  values of 
oscillatory energy of impurity atoms  olso brought, calculated on the
eq.~(\ref{harm}) with provision for values ($E_0 = 0.283 eV$) for Al and
($E_0 = 0.301 eV$) for~S. Modification of optical reflection~(dR) 
caused by impurity atoms Al in GaP are presented on fig.~3. Experimental 
temperature dependencies of TEP for GaP samples are presented on fig.~4. 
Temperature dependencies of TEP for samples Si(P,O) and for polisilicon
film on the silicon substrate (Si$^*$) are presented on fig~5.

Cuve~1 on fig.~2 may be characterized with activation energy 
$E_a$~$\cong$~0.7~meV at temperatures T$<$330K but value $E_a$ is close 
to the width of GaP forbidden gape (2.4 eV) at temperatures $T>330K$.
Curve~2 and~3 on fig.~2  may be characterized by several activation 
energies. Corresponding values $E_a$ are contained in Table~1 and Table~2
and may be explained differently. Values $E_a$ in lines of Table~1, 
marked by the stars in the first column, close to  crystalline phonons which
intensively interact with local centers. These phonons are pointed out in 
central column in Table~1. It is impossible to explain the experimental 
dependencies $\rho$(T) by dissipation of charge carriers by phonons since the 
dissipation is capable to create an opposite effect to the observed 
effect of reducing resistance when increasing the temperature. We connect 
these energies with generation of free charge carriers from 
electron-vibrational levels of local centers  created by atoms Al or~S. 
Other energies in Table~1 supposedly  may be explained by generation of free 
charge carriers from electron-vibrational levels formed  by 
inherent oscillations ($\beta$ and $\gamma$-types) of Al or S and 
possible by combinations of such oscillations with cristalline phonons.

\begin{table*}[t]
\begin{center}
\caption{Activation energies for GaP(Al) and GaP(S) at $T<330K$}
\vspace{0.5cm}
\begin{tabular}{|c|c|c|c|c|c|c|c|c|c|c|}
\hline
\multicolumn{5}{|c|}{Activation energies $E_a$ (meV) for} & Phonons in &
\multicolumn{5}{|c|}{Activation energies $E_a$ (meV) for }\\
\multicolumn{5}{|c|}{GaP(Al) samples with numbers:} & GaP (meV)  & 
\multicolumn{5}{|c|}{GaP(S) samples with numbers:}\\
\cline{1-5} \cline{7-11} 
\ 1 & 2 & 3 & 4 & 5 & \cite{Mar64} & 6 & 7 &8 & 9 & 10\\
\hline
\ ~~~8.1 & ---& --- & 7.2 & 8.0 & --- & --- & --- & --- & --- & ---\\
\hline
\ *~~15.0 & --- & --- & 14.8 & 14.5 & 14.25 & 15.0 & 14.8 & --- & --- & 14.3\\
\hline
\ *~~24.5 & --- & --- & --- & 25.0 & 24.42 & 24.0 & --- & 24.6 & --- & ---\\
\hline
\ --- & 35.0 & --- & --- & --- & --- & --- & --- & --- & 28.0 & 28.0\\
\hline
\ *~~45.0 & --- & --- & --- & 44.6 & 44.75 & --- & 42.0 & --- & --- & ---\\
\hline
\ *~~---~~~ & --- & 49.0 & 48.0 & --- & 47.00 & 47.0 & --- & --- & --- & ---\\
\hline
\ ~~~75.0 & ---& 83.0 & --- & --- & --- & --- & --- & 70.0 & --- & ---\\
\hline
\end{tabular}
\end{center}
\end{table*}

\begin{table*}[t]
\begin{center}
\caption{Activation energies for GaP(Al) and GaP(S) at $T>330K$}
\vspace{0.5cm}
\begin{tabular}{|c|c|c|c|c|c|c|}
\hline
\multicolumn{5}{|c|}{$E_a$ (eV) for GaP(Al)} & Calculated & Multiple $E_0$\\
\multicolumn{5}{|c|}{samples whith numbers:} & on eq. (\ref{harm}) & (eV)\\
\hline
\ 1 & 2 & 3 & 4 & 5 & --- & ---\\
\hline
\ 0.14 & 0.14 & 0.14 & 0.138 & 0.137 & 0.1415 ($\nu=0$) & ---\\
\hline
\ 0.28 & 0.29 & 0.29 & 0.28 & 0.28 & --- & $E_0=0.283$\\
\hline
\ 0.42 & 0.42 & 0.43 & 0.42 & 0.43 & 0.4245 ($\nu=1)$ & ---\\
\hline
\ --- & 0.57 & --- & 0.56 & 0.58 & --- & $2E_0=0.566$\\
\hline
\ 0.71 & --- & 0.72 & --- & --- & 0.7075 ($\nu=2$) & ---\\
\hline
\ --- & --- & --- & 0.85 & --- & --- & $3E_0=0.849$\\
\hline
\ --- & 0.97 & --- & --- & --- & 0.9905 ($\nu=3$) & ---\\
\hline
\ 1.10 & --- & --- & --- & 1.11 & --- & $4E_0=1.132$\\

\hline

\multicolumn{5}{|c|}{$E_a$ (eV) for GaP(S)} & --- & ---\\
\multicolumn{5}{|c|}{samples whith numbers:} & --- & ---\\
\hline
\ 6 & 7 & 8 & 9 & 10 & --- & ---\\
\hline
\ 0.15 & 0.15 & 0.15 & 0.15 & 0.15 & 0.1505 ($\nu=0$) & ---\\
\hline
\ 0.3 & 0.29 & 0.3 & 0.3 & 0.31 & --- & $E_0=0.301$\\
\hline
\ --- & --- & --- &  --- & --- & 0.4515 ($\nu=1$) & ---\\
\hline
\ 0.6 & --- & 0.6 & 0.61 & --- & --- & $2E_0=0.602$\\
\hline
\ --- & --- & --- & --- & 0.74 & 0.7525 ($\nu=2$) & ---\\
\hline
\ -- & --- & 0.92 & --- & --- & --- & $3E_0=0.903$\\
\hline
\ 1.03 & --- & --- & --- & --- & 1.0500 ($\nu=3$) & ---\\
\hline
\end{tabular}
\end{center}
\end{table*}

It is seen from Table~2 that activation energies of samples with
each type of impurities can be divided into two groups which
corresponds to two  columns dextral. One group consists of activation 
energies which are described by the equation for quantum
harmonic oscillator~(\ref{harm}). These activation energies are 
connected with inherent oscillations of impurity atoms and correspond 
to transitions from oscillatory energy levels with~$\nu=0,~1,~ 2,~\cdots$ 
in the minimum of oscillatory potential where oscillatory energy is zero. 
Such transition for free quvantum harmonic oscillator are forbidden but 
they are possible for classical oscillator. Consequently, inherent 
oscillators of impurity atoms show duality of properties that can be 
explained by their interaction with the cristal. Other group of energies 
in Table~2 consists of energies multiple $E_0$. Energies of 
this group are also connected with inherent oscillations of impurity 
atoms and correspond to transitions between different oscillatory 
energy levels (between levels with different $\nu$). It is seen that 
the value $E_0$ is the same for both groups of energies. Consequently, 
both groups of energies refer to  the same type of centres 
that show a classical and quantum properties (duality of properties).

Studies of infra-red reflection (R) also confirm presence of inherent 
oscillations in GaP(Al) and GaP(S) samples  and their intensive 
interaction with electrons and crystalline phonons. Modification of 
infra-red reflection spectrum~(dR) caused by impurities atoms was separated 
on  components in accodance with complex oscillatory  
model~\cite{Ros51,Nozi58} taking into account the contributions from j 
different charged oscillators. 
Theese components  are marked on fig.~3 by numerals j:~2,~3,~4 and~5. 
Every component involved contribution into reflection~(R)  reach the 
highest value when optical frequency ($\omega$) satisfy to the conditions 
$\omega_p>\omega>\Omega$.
$\Omega$ - an oscillator frequency and $\omega_p$ - a frequency of 
springy oscillations of lattice. Minimum~dR locates near $\omega_p$.  
Agreement of experimental~-~1 and amount of calculated spectrums~2,~3,~ 4 
and~5 is reached if energies $\hbar\omega_p$ are equal to the energies of 
$\alpha$~-~tipe inherent oscillations for Al:~0.5$E_0$,~$E_0$, 
$1.5E_0$,~$2E_0$. Two from them comply with
calculated on eq.~(\ref{harm}) if $E_0=0.283$~eV and $\nu=0$ or~1 but 
two other are multiple the same~$E_0=0.283$~eV. Reflection spectrums 
of GaP(S) also are  described within the complex osacillator 
model~\cite{Ros51,Nozi58} if~$E_0=0.301$~eV. Energies~$\hbar\Omega$ for 
both types of impurities~(Al,S) may be connected with $\gamma$~-~type of 
inherent oscillations Al(61.1 meV) or S(65.0 meV). Energy loss of 
inherent oscillators is great (G/$\Omega\simeq$0.09; G- damping factor) 
and corresponds 
to the tight binding of impurity atoms with the cristalline phonons. 
Thereby centres formed  by impurities Al and S in GaP show duality of
optical properties. Theese properties are defined by interaction of
inherent oscillations of impurities atoms, crystalline phonons and 
electrons with  each other. 

The best consent of calculated and experimental reflection spectrums 
is reached if optical dielectric permeability~($\varepsilon\cong2$)
is small in contrast with the highfrequency dielectric permeability 
GaP~($\varepsilon=8.457$)~\cite{Klei60}. Probably measured value
of dielectric permeability follow refere to local centres in which 
electronic optical transitions occur.
A cross section of photons capture  can be defined by wavelength of 
cristalline phonons which interact with the centres. 

Inherent oscillations can spread in crystals as waves
of inherent oscillations. Inherent oscillations, crystalline phonons 
and electrons~(holes) powerfully interact with each other and 
give rise to electrical currents. This is confirmed by particularities 
of TEP. Curve~1 on fig.~4 is typical 
temperature dependency of TEP for samples GaP(S). Curve~2 on fig.~4 is 
typical for GaP samples without impurities and may be explained by close 
to inherent conductivity. Temperature dependency of TEP for GaP(Al) is 
like curve~1. Curve~1 on fig.~4 keeps pius which marked  by arrows and by 
letters. Polarity of these  pius agrees with the polarity of drift TEP.
We connect these  pius  with the phonon drag of electrons. Pius~A,~B,~C
and~F are situated at Debye temperatures of cristalline phonons in 
GaP~\cite{Mar64}: 95K~(TA,8.2~meV); 168K~(TA,~14.25~meV); 
288K~(LA, 24.42~meV); 542K~(LO,~44.75~meV). Broad pius~D~($\approx345K$) 
and E~($\approx475K$) can to be explained by combinations of cristalline 
phonons: 
(TA+TA,~28.6~meV) and (TA+LA,~38.67~meV). Investigations of $\rho$(T), 
TEP temperature dependency and infra-red reflection show that inherent 
oscillations Al and S in GaP actively interact with cristalline phonons 
and electrons.
This ensures electron-phonon interaction sufficient for the realization 
of phonon drag effect (and  probably superconductivity) at very high 
temperatures.

Curve~1 on fig.~5 represents peculiar temperature dependencies of TEP for
samples Si(P,O). It has pius spesified by arrows and by 
letters~a,~b,~c,~d,~e. Polarity of these pius coincides with the polarity of 
drift TEP. 
Pius~a,~b,~c are situated at Debye temperatures of acoustic phonons 
with wave vektors along certain directions~\cite{Aubr63}: 
$<$111$>$,~200.4K~(16.7~meV); $<$110$>$,~214.8K~(17.9~meV); 
$<$100$>$,~252K~(21.0~meV). We explain these pius  by phonon drag of electrons.
Pius d and e we involve with TO phonons in~Si. Curve~2 on fig.~5 
represents the temperature dependency of TEP in Si$^*$ and has 
pius~p,~q,~r,~h with different polarities. Temperatures of these pius 
are equal to Debye temperatures of phonons in critical points in the
Brillouin area of Si: L(W)~-~551K~(45.9~meV); L(L)~-~606K~(50.5~meV);
TO(X)~-~683K~(56.9~meV) and TO(L)~-~712K~(60.9~meV) accodingly.
We explain pius~f,~g,~h,~i  by phonon drag of holes but peak j we 
explain by phonon drag of electrons.

The sufficient interaction between electrons (houls) 
and phonons in silicon samples can be explained by inherent 
oscillations of centres which are formed by impurities of Oxygen 
($E_0$=0.22~eV). These are to be A-centres. Such interpretation agrees
with  experimental temperature dependencies of electrical resistivity 
for the same samples. These dependencies are discribed by several activation 
energies among which there are multiple $E_0=0.22$~eV or equal to 
calculated on the eq.~\ref{harm} if $E_0$ is equal to the same value.

\section*{5.~'onclusion}
\hspace{1.5em}
 
The analysis of experimental results has shown that phonon drag effect is
caused by the phonons which actively interact with electron-vibrational 
centres formed by impurity atoms (Sulphur and Aluminium in GaP, Oxygen 
in silicon samples). Such centres can generate  inherent oscillations 
and waves. Theoretical possibility of strong electron-phonon interaction, 
caused by inherent oscillations and waves, was confirmed 
by the phonon drag effect at temperatures up to about 700K. 
The phonon drag component of thermoelectric power also 
has confirmed  the presence of inherent oscillations waves in crystals.
The expected  superconductivity from about room temperatures up to 
temperatures more~700K  provided by inherent oscillations and waves 
probably will be realized.

\vspace{0.5cm}

\newpage

\section *{Figure captions}

\newcommand{\Fig}[2]{\noindent{\bf Fig.~#1.~}
\parbox[t]{15cm}{\baselineskip 24pt #2}\\[5mm]}

\Fig{1}
{Inherent~(I) and acoustic~(A) branches of oscillations in adiabatic 
simple univariate crystal model ( 1~-~at $\delta$=0; 2~-~at $\delta>$0;
3~-~at $\delta<$0 ). Fragment of adiabatic simple univariate crystal 
model is shown on insertion.}

\Fig{2}
{Temperature depenencies of electrical resistivity $\rho$(T) for samples 
GaP: 1~-~without impurities, 2~-~with impurity of Aluminium; 3~-~ with 
impurity of Sulphur.}

\Fig{3}
{Change of optical reflection (dR) caused by impurity of Aluminium 
in GaP$~-~$1. Calculated components of dR connected with inherent 
oscillations of Aluminium~:~2,~3,~4,~5.}

\Fig{4}
{Temperature dependency of thermoelectric power: 1~-~in GaP with impurity 
of Sulphur; 2~-~in GaP withaut impurities.}

\Fig{5}
{Temperature dependency of thermoelectric power: 1~-~in Si with impurity of 
Phosphorus and Oxygen; 2~-~in porous silicon ($Si^*$).}
 
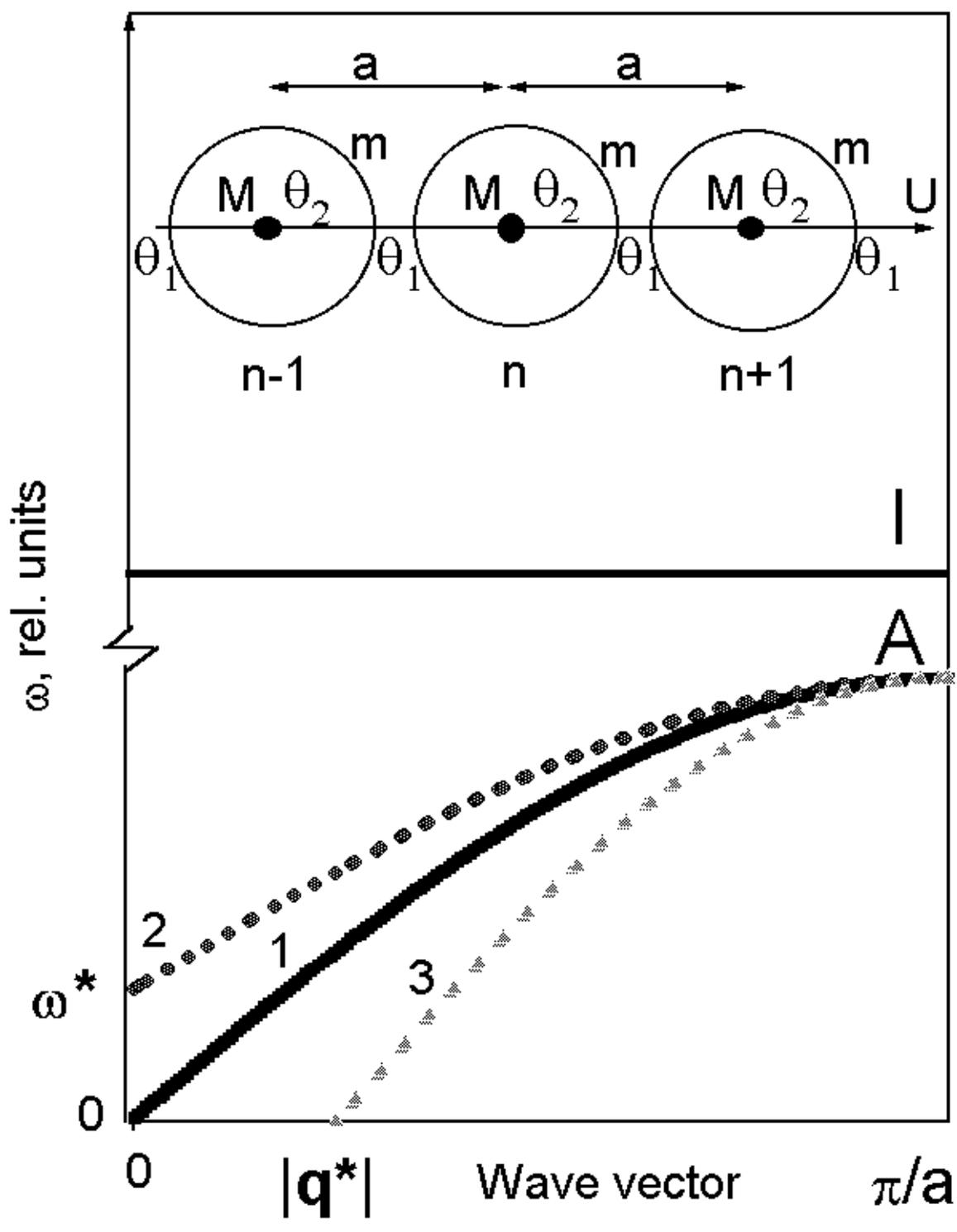
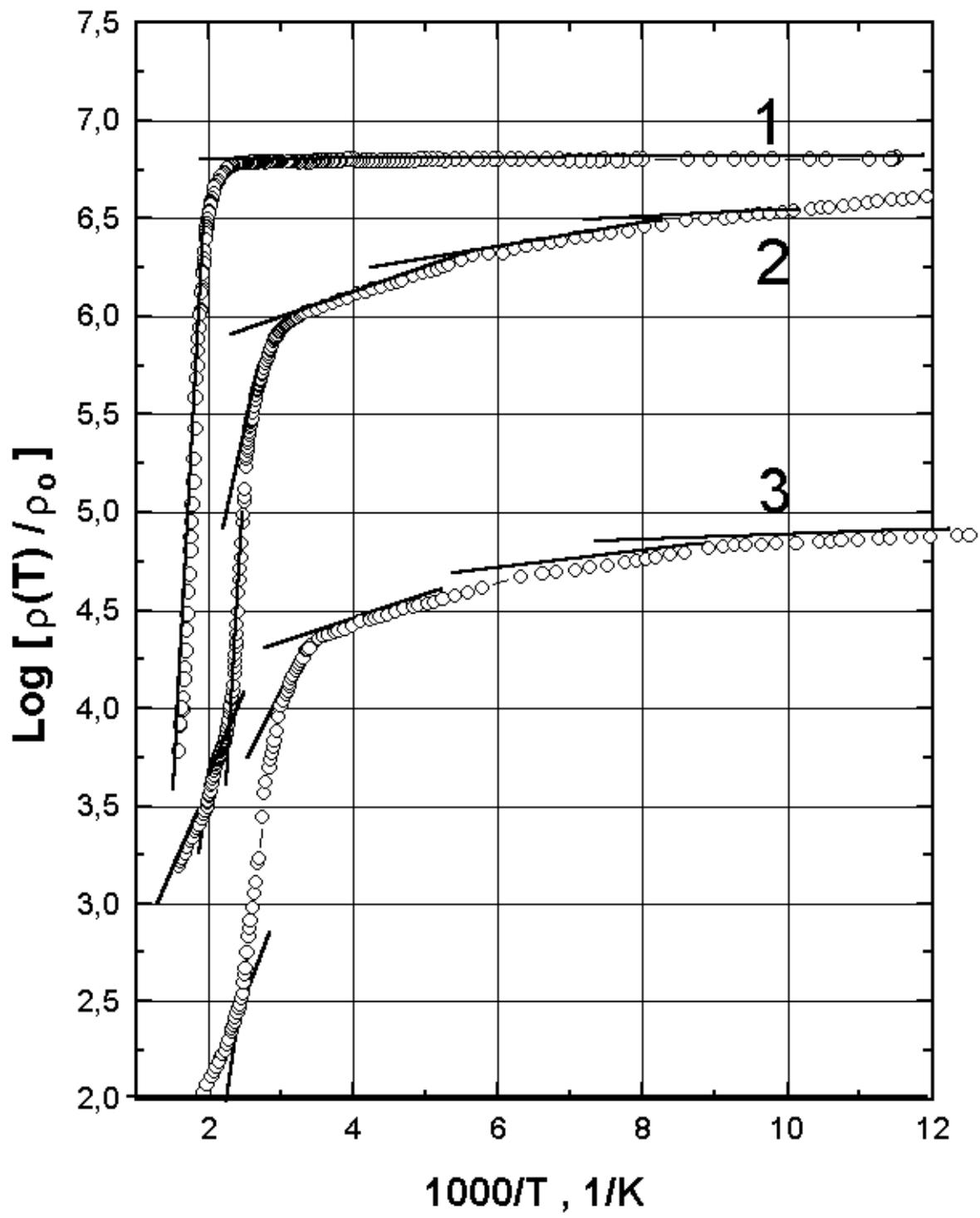
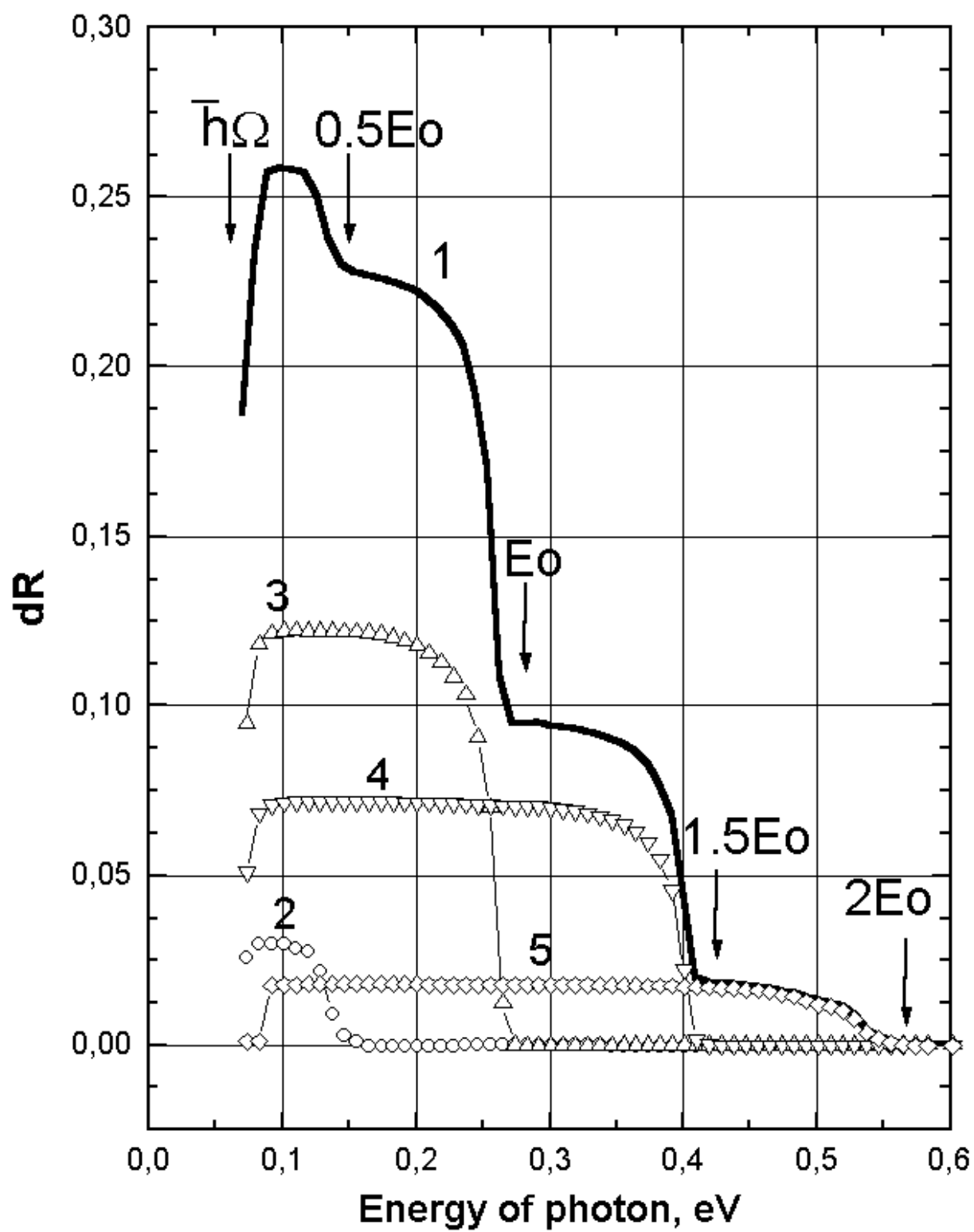
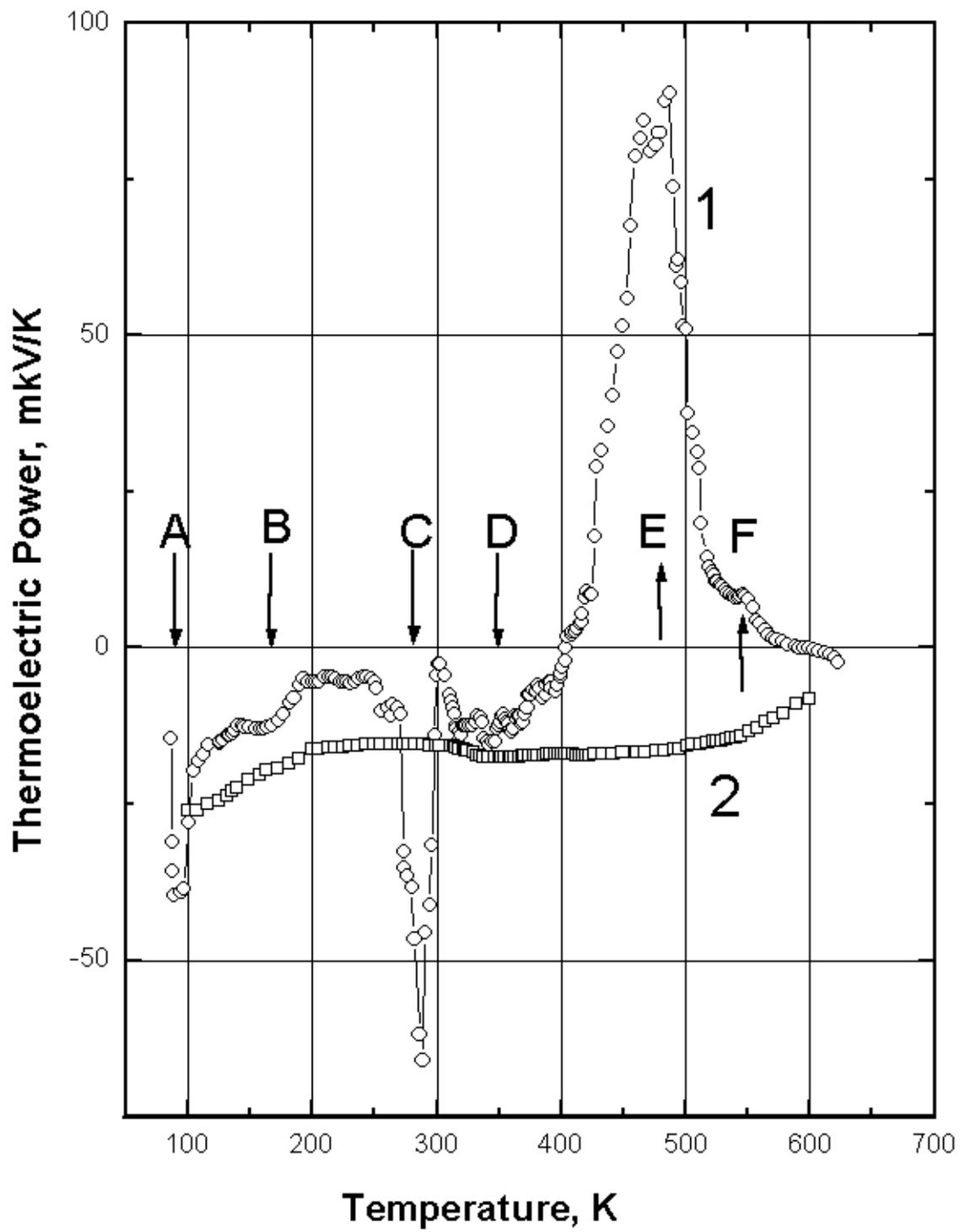
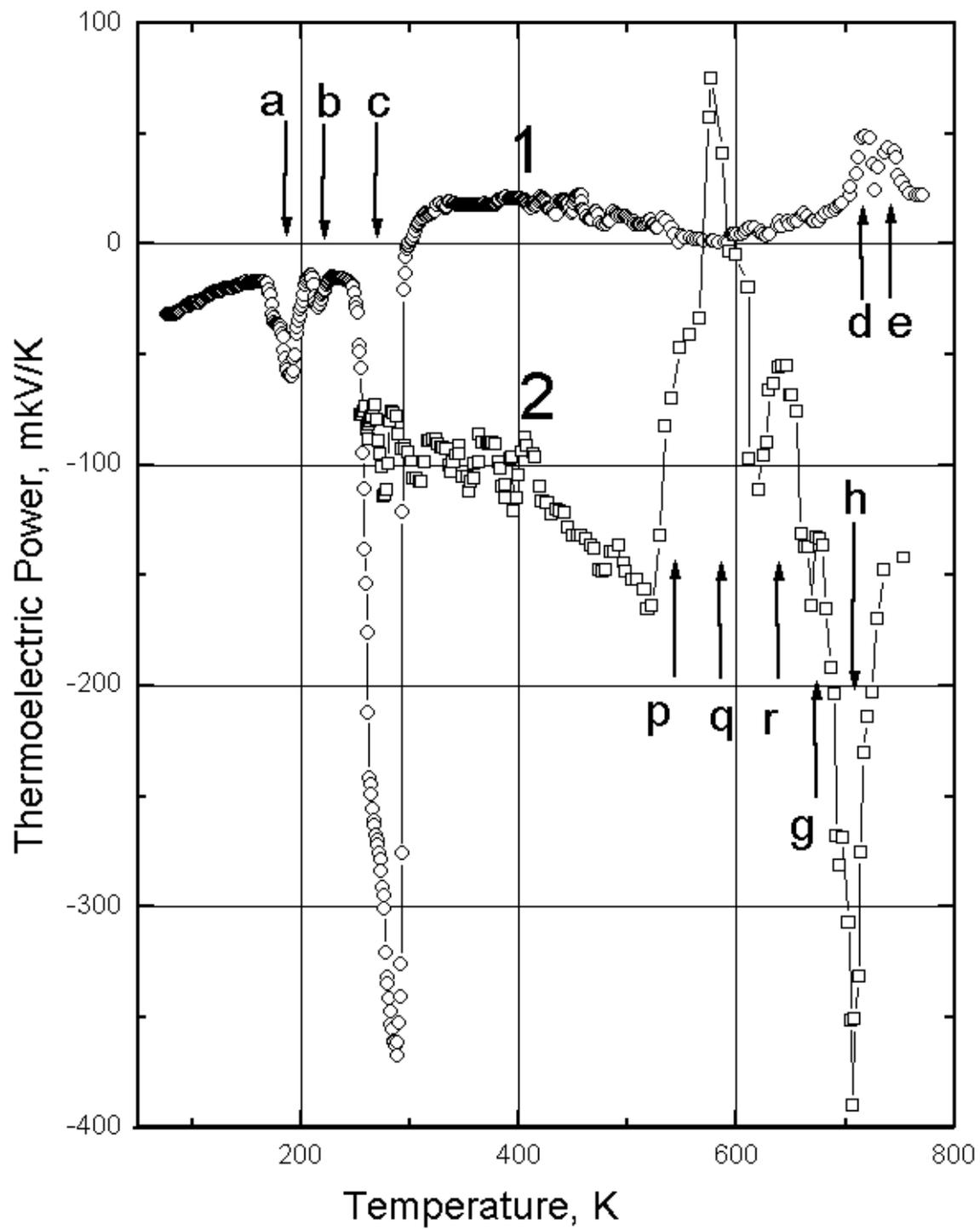
\end{document}